\documentclass[]{aastex631}

\shorttitle{Supra decadal modulation in solar activity}
\usepackage{mathtools}

\graphicspath{{./}{figures/}}

\begin{document}

\title{On the origin of long-term modulation in the Sun's magnetic activity cycle}

\author[0000-0002-3131-8260]{Chitradeep Saha}
\affiliation{Center of Excellence in Space Sciences India, 
Indian Institute of Science Education and Research Kolkata,\\
Mohanpur 741246, West Bengal, India }

\author[0009-0008-0995-7120]{Suprabha Mukhopadhyay}
\affiliation{Center of Excellence in Space Sciences India, 
Indian Institute of Science Education and Research Kolkata,\\
Mohanpur 741246, West Bengal, India }
\affiliation{Department of Physical Sciences,   
Indian Institute of Science Education and Research Kolkata,\\
Mohanpur 741246, West Bengal, India}
\affiliation{Max-Planck-Institut f{\"u}r Sonnensystemforschung, Justus-Von-Liebig-Weg 3, 37077 G\"ottingen, Germany}

\author[0000-0001-5205-2302 ]{Dibyendu Nandy}
\affiliation{Center of Excellence in Space Sciences India, 
Indian Institute of Science Education and Research Kolkata,\\
Mohanpur 741246, West Bengal, India }
\affiliation{Department of Physical Sciences,   
Indian Institute of Science Education and Research Kolkata,\\
Mohanpur 741246, West Bengal, India}

\begin{abstract}

One of the most striking manifestations of orderly behavior emerging out of complex interactions in any astrophysical system is the 11-year cycle of sunspots. However, direct sunspot observations and reconstructions of long-term solar activity clearly exhibit amplitude fluctuations beyond the decadal timescale -- which may be termed as supradecadal modulation. Whether this long-term modulation in the Sun’s magnetic activity results from nonlinear mechanisms or stochastic perturbations remains controversial and a matter of active debate. Utilizing multi-millennial scale kinematic dynamo simulations based on the Babcock-Leighton paradigm -- in the likely (near-critical) regime of operation of the solar dynamo -- we demonstrate that this supradecadal modulation in solar activity cannot be explained by nonlinear mechanisms alone; stochastic forcing is essential for the manifestation of observed long-term fluctuations in the near-critical dynamo regime. Our findings substantiate some independent observational and theoretical investigations, and provide additional insights into temporal dynamics associated with a plethora of natural phenomena in astronomy and planetary systems arising from weakly nonlinear, non-deterministic processes.
 
\end{abstract}

\section{Introduction} \label{sec:intro}
The dynamic activity of our host star, the Sun, manifests itself across a wide range of timescales. The typical 11-year solar cycle -- characterized by the almost regular surge and ebb in the number of sunspots, i.e. strongly magnetized dark patches on the Sun's surface -- is a well-studied phenomenon \citep{HATHAWAY2015LIVREV}.  Long-term reconstruction based on cosmogenic isotopes also reveals the existence of quasiperiodicities at much higher timescales \cite[][and references therein]{SONETT19990PTRSL,  MIYAHARA2004SOLPHYS, USOSKIN2023LIVREV}. The origin of this `supradecadal' modulation and the manifestation of these quasi-periodicities are poorly understood. Nevertheless, it is universally acknowledged that supradecadal modulation of the solar activity cycle is a significant constraint on underlying physical processes that sustain the magnetohydrodynamic (MHD) solar dynamo mechanism. Understanding the basis of these fluctuations is critical to revealing how complex physical processes in convection zones lead to the origin and variability of magnetism in solar-like stars.

Solar magnetic activity cycles are deep rooted in the Sun's convection zone, wherein the inductive action of the large-scale MHD dynamo mechanism periodically generates and recycles magnetic fields and sustains them against Ohmic dissipation \citep{CHARBONNEAU2020LIVREV}. Understanding and predicting solar variability across diverse timescales \citep{Hazra2023SSR, Bhowmik2023SSR} is crucial as solar radiation, magnetic flux, and particulate output modulate planetary space environment and atmospheres, thereby determining space weather and habitability \citep{BRAUN2005NATURE, SCHRIJVER2015AdvSR, NANDY2023JASTP, Gupta2023ApJ}.

Many studies in the past have noted quasiperiodicities extending over multiple sunspot cycles, including the centennial Gleissberg cycle, $\sim$200-year Suess/de Vries cycle, $\sim$240-year Hallstatt cycle, and other long-term modulations traced from cosmogenic isotope proxies \citep{VASILIEV2002ANNALES, SCAFETTA2016ESR, USOSKIN2016AANDA, BEER2018MNRAS}. In spite of being the subject of intense scrutiny, the physical origin of this supradecadal modulation in solar activity is not yet well understood; specifically, the primary debate centers around whether nonlinear mechanisms or stochastic perturbations create supradecadal modulation in the solar cycle.  

On the one hand, dynamical (deterministic) chaos arising from nonlinear feedback involved in the solar dynamo mechanism is believed to play a key role in driving super-secular solar cycles  \cite[see,][]{USOSKIN2023LIVREV, BISWAS2023ISSI}. In this approach, the Sun is often considered an oscillator, and its dynamics is modeled by a set of nonlinear equations with reduced dimensionality that can exhibit chaotic behavior in the suitable regime of parameter space \cite[][and references therein]{WEISS2016MNRAS}. As pointed out in \cite{PANCHEV2007JASTP} existing nonlinear time series analysis methods are prone to obtain spurious indications of low-dimensional dynamical chaos in the solar activity record, making the task challenging.  

On the other hand, several studies have identified stochastic perturbations associated with turbulent plasma motions in the solar convection zone (SCZ) as an essential ingredient in generating supradecadal modulation \citep{MININNI2000PRL, MININNI2002PRL}.  Considerable dispersion around the mean tilt angle of bipolar sunspot pairs as gleaned from solar photospheric observation is understood to be due to turbulent buffeting (i.e., stochastic perturbation) of magnetic flux tubes rising through the solar convection zone. This eventually introduces a random component in the solar polar field build-up, thereby causing modulation in cycle amplitudes in the Babcock-Leighton (BL) paradigm \citep{Choudhuri1992AandA, Charbonneau2000ApJ, DASI2010AANDA, CAMERON2015SCIENCE, NAGY2017SolPhys, PAL2023APJ}. Further sources of perturbation could be red noise generated from solar surface turbulence \citep{MATTHAEUS2007APJ, NICOL2009APJ} and an independent, small-scale turbulent dynamo that acts as a source of ``magnetic noise'' \citep{Rempel2023SSR}.

Nonlinear mechanisms with varying intensities in the highly supercritical regime can potentially generate a wide range of dynamical behavior in dynamo solutions, regardless of the level of stochastic forcing \citep{CAMERON2017APJ, BUSHBY2006MNRAS}. However, it is crucial to note that recent stellar gyrochronology observations, complemented by numerical simulations, strongly indicate that the solar dynamo is currently operating in a sub-critical or near-critical regime \citep{Metcalfe2016ApJL, vanSaders2016Nature, REINHOLD2020SCIENCE, TRIPATHI2021MNRASL} -- implying that the solar dynamo is only weakly nonlinear. It is, therefore, imperative in this context to understand the relative roles of stochastic perturbation and nonlinear (chaotic) feedback on the dynamics of solar magnetic cycles over longe timescales in this near-critical regime.

Simulations using a diversity of dynamo models have proven to be sufficiently potent in understanding long-term solar magnetic variability in greater detail \citep{KARAK2023LIVREV}. 

In this study, we utilize stochastically driven numerical solar dynamo models with nonlinear algebraic quenching in the poloidal sources to shed light on the interplay of stochasticity and nonlinearity in generation of long-term solar activity modulation beyond the decadal timescale in the near critical regime of dynamo operation. Our simulations show that invoking nonlinear quenching mechanisms alone -- in the absence of any stochastic forcing in the dynamo models we explore -- is unable to produce statistically significant and persistent supradecadal modulation in solar magnetic activity.

\section{Methods \& RESULTS} \label{sec:setup}

\begin{figure*}
    \begin{center}
    \includegraphics[width=0.8\linewidth]{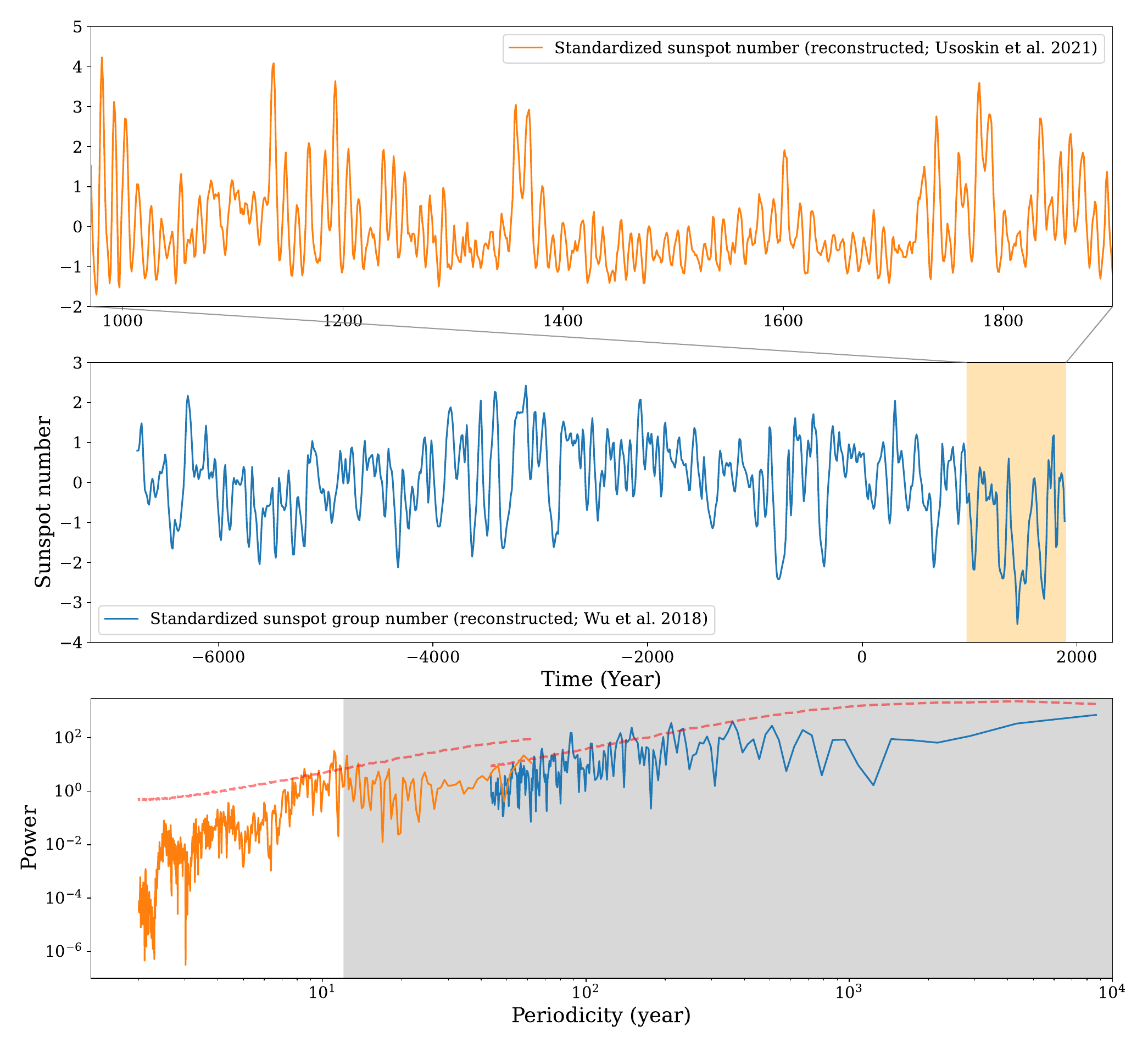}
    \caption{Top and middle panels: plot of reconstructed decadal sunspot group number \citep[][blue curve]{WU2018VIZIER}, with annually resolved reconstructed sunspot number series \citep[][orange curve]{USOSKIN2021AANDA}.      Bottom panel: power spectral density corresponding to the above time series.  In each cases, the red-dashed curves mark the upper threshold of the 3-$\sigma$ significance level of power spectra generated out of 5,000 noise-based surrogate realizations of the original data. We estimate the first order auto-regressive parameters (AR1) \citep{ALLEN_1996_JSTOR} to model the colored noise in the time series. Auto-regressive model of first order should be sufficiently good over higher orders in this context as we are mainly probing the longer timescale regime where the auto-correlation in the time series becomes weaker. The gray-shaded region signifies the periodicity interval that corresponds to amplitude modulation beyond decadal timescale, i.e., supradecadal modulation.}
    \label{fig:dat1}
    \end{center}
\end{figure*}

For the present study, we rely on numerical solar dynamo modeling and compare the simulated results with observed reconstructions of solar activity \citep{WU2018VIZIER} to validate our findings. Particularly, we use the reconstructed sunspot number time series which captures well the long-term modulation in solar cycles in the past nine millennia (see Fig.\ref{fig:dat1}). To simulate solar magnetic cycles, we employ a spatially extended two-dimensional and a dimensionally reduced dynamo model -- both of which have been extensively used and benchmarked in earlier studies \citep[see, e.g.,][]{NANDY2002SCIENCE, CHATTERJEE2004AANDA, PASSOS2014AANDA, TRIPATHI2021MNRASL, SAHA2022MNRASL}.  The global magnetic field of the Sun, $\mathbf{B}$, can be expressed as a combination of a toroidal component B$_{\phi}$$\mathbf{e}_{\phi}$  and a poloidal component $\mathbf{B_p}$ in the spherical coordinate system. Both models solve for the time evolution of these two components in the form of coupled partial differential equations.

The spatially extended model works in the kinematic regime with an axisymmetry assumed around the solar rotation axis. Large-scale plasma flows are completely specified on the meridional plane and are kept fixed throughout the simulation domain. Additional source terms are introduced which govern the generation of poloidal field, $\mathbf{B_p}$ (i.e., $A$) from toroidal component, $B_{\phi}$, as described in the following Eqs.\eqref{eq:s1} and \eqref{eq:s2},

\begin{equation}\label{eq:s1}
\frac{\partial A}{\partial t}+\frac{1}{s}(\textbf{v}_p \cdot \nabla)(sA)=\eta_p \left(\nabla^2-\frac{1}{s^2}\right)A+\alpha B_{\phi}
\end{equation}
\begin{multline}\label{eq:s2}
\frac{\partial B_{\phi}}{\partial t}+s\left[ \textbf{v}_p \cdot \nabla \left( \frac{B_{\phi}}{s}\right) \right] + (\nabla\cdot \textbf{v}_p) B_{\phi} =\eta_t \left(\nabla^2-\frac{1}{s^2}\right)B_{\phi}
+s(\textbf{B}_p.\nabla)\Omega +\frac{1}{r}\frac{d \eta_t}{dr}\frac{\partial }{\partial r} (rB_{\phi})
\end{multline}

\noindent
where, $s=r \sin \theta$ . The magnetic vector potential, $A\textbf{e}_\phi$, is associated with $\textbf{B}_p$ through the mathematical relation, 
$\textbf{B}_p(r,\theta ,t)=\mathbf{\nabla} \times [A(r,\theta ,t)\textbf{e}_{\phi}] $. Meridional plasma flow is denoted by $\mathbf{v_p
}$, whereas $\Omega$ represents the rotation in the solar convection zone. We use a peak flow speed of 29 ms$^{-1}$ for the meridional circulation. The symbols $\eta_p$, and $\eta_t$ denote the poloidal and toroidal diffusivity, with magnitude of  2.6$\times$10$^{12}$  cm$^2$s$^{-1}$ and 4$\times$10$^{10}$  cm$^2$s$^{-1}$ used in our simulations, respectively \cite[similar set of parameters as that used by][to generate their Fig. 6]{PASSOS2014AANDA}.
The poloidal source $\alpha$ comprises two different terms -- the mean-field $\alpha_{MF}$ that acts throughout the convection zone, and  $\alpha_{BL}$ that operates near the surface to emulate the Babcock-Leighton (BL) mechanism. Mathematical parametrization of these quantities along with their justifications are available in earlier studies \cite[see, e.g.,][]{CHATTERJEE2004AANDA, YEATES2008APJ, PASSOS2014AANDA}.

To simulate long-term fluctuations in the solar dynamo, various modeling approaches exist which invoke non-linear mechanisms \citep{WILMOT2005MNRAS, JOUVE2010AAP, WEISS2016MNRAS}, or a combination of non-linearity and  stochasticity in the process of poloidal field generation, or fluctuation in the meridional circulation \citep{CHOUDHURI2012PRL, Olemskoy2013ApJ, HAZRA2014APJ, PASSOS2014AANDA, Karak2017APJ, HAZRA2019MNRAS}. In the context of the emerging understanding that the BL mechanism for poloidal field generation is the primary driver of solar cycle variability, nonlinearities in the BL poloidal source such as tilt quenching \citep{DSILVA1993AANDA, JHA2020APJL} and latitude quenching \citep{Jiang2020ApJ, Karak_2020_ApJl_latq} have been proposed, observed and quantified \citep{Yeates2025ApJ}. In our model, we use algebraic $\alpha$-quenching of poloidal sources which captures some essential aspects of these observed nonlinear physical mechanisms in a generic way. The poloidal source terms scale with the toroidal field strength in a nonlinear fashion as shown below,

\begin{align}\label{eq:q1}
    \alpha_{MF} \sim  \left({1+\left(\dfrac{B_{\phi}}{B_{eq}}\right)^2}\right)^{-1}&\\ \label{eq:q2}
    \alpha_{BL} \sim  \left(1+\mathrm{erf}{\left(\frac{B_{\phi}^2-B_{lo}^2}{d_{lo}^2}\right)}\right)  &\left(1-\mathrm{erf}{\left(\frac{B_{\phi}^2-B_{up}^2}{d_{up}^2}\right)}\right)
\end{align}

\noindent
In Eq.\eqref{eq:q1}, $B_{eq}$ (=10 kG) denotes the equipartition field strength -- assuming an equipartition between the magnetic and turbulent energies in the solar convection zone -- above which magnetic back reaction on large-scale flows becomes significant. The formulation for $\alpha_{BL}$ in Eq.\eqref{eq:q2} ensures that only the fields that reach the surface layers with magnitudes between $B_{lo}$ (=1 kG) and $B_{up}$ (=100 kG) contribute to the BL mechanism in our simulations. The motivation behind implementing such thresholds is  that the weak toroidal flux tubes are shredded by turbulence and are unable to form sunspots, whereas very strong flux tubes rise rapidly, which causes them to emerge at the solar surface with no significant
tilt -- thereby quenching the BL mechanism. Values of remaining aforementioned parameters are adapted from \cite{PASSOS2014AANDA}.

In a number of our present simulations, we introduce stochastic fluctuation in poloidal sources (i.e., $\alpha_{BL}$ and $\alpha_{MF}$) to mimic the scatter in the tilt of bipolar magnetic regions around the Coriolis force-induced mean tilt angle \citep{DASI2010AANDA}.  Mathematically, the fluctuation is modeled as,
\begin{equation}\label{eq:fl}
    \alpha =\alpha^0\left[ 1+\delta\sigma(t,\tau)\right]
\end{equation}
where,  $\alpha^0_{BL}$=27 ms$^{-1}$ and $\alpha^0_{MF}$=0.4 ms$^{-1}$ represent the mean amplitudes of $\alpha_{BL}$ and $\alpha_{MF}$, respectively. Around these mean values, time dependent uniform white noise $\sigma (t,\tau)$ with an amplitude $\delta$ and a coherence time $\tau$ is imparted to the system. The amplitude $\delta$ takes the value of 150$\%$ for $\alpha_{BL}$ and 100$\%$ for $\alpha_{MF}$. The coherence time, $\tau$, is set to 6 months for $\alpha_{BL}$ and 1 month for $\alpha_{MF}$.

To perform a comparative assessment and establish the robustness of our finding, we also perform solar cycle simulations with an independent and well-studied low-order dynamo model based on stochastically forced delay differential equation \citep{WILMOT2006APJ, HAZRA2014APJ, TRIPATHI2021MNRASL}. Similar to the spatially extended model, the reduced model also imbibes the essence of BL mechanism in the source term $\alpha_{BL}$ as follows,

\begin{equation}
\frac{\mathrm{d}A(t)}{\mathrm{d}t}=\alpha_{BL} \cdot f\left(B_{\phi}(t-T_1)\right) B_{\phi}(t-T_1)-\frac{A(t)}{\tau}+ \epsilon(t)
\end{equation}

\begin{equation}
 \frac{\mathrm{d} B_{\phi}(t)}{\mathrm{d}t}=\frac{\omega}{L} A(t-T_0)-\frac{B_{\phi}(t)}{\tau}
 \end{equation}

\noindent
The form of nonlinear quenching and stochastic fluctuation in $\alpha_{BL}$ employed in this model is similar to Eqs.\eqref{eq:q2} and \eqref{eq:fl}, respectively, except that the amplitude of fluctuation in the poloidal source, $\alpha_{BL}$, in this case is 30$\%$ and for the additive magnetic noise, $\epsilon$, it is 5$\%$. To clarify further, the amplitude of stochastic fluctuations in all our simulations are set to closely match with the occurrence statistics of grand minima gleaned from the long-term reconstruction of solar magnetic activity \citep{USOSKIN2023LIVREV}. An elaborate description of the stochastic time delay dynamo model and the values of other parameters can be found in \cite{TRIPATHI2021MNRASL}.

\begin{figure*}
    \begin{center}
    \includegraphics[width=0.95\linewidth]{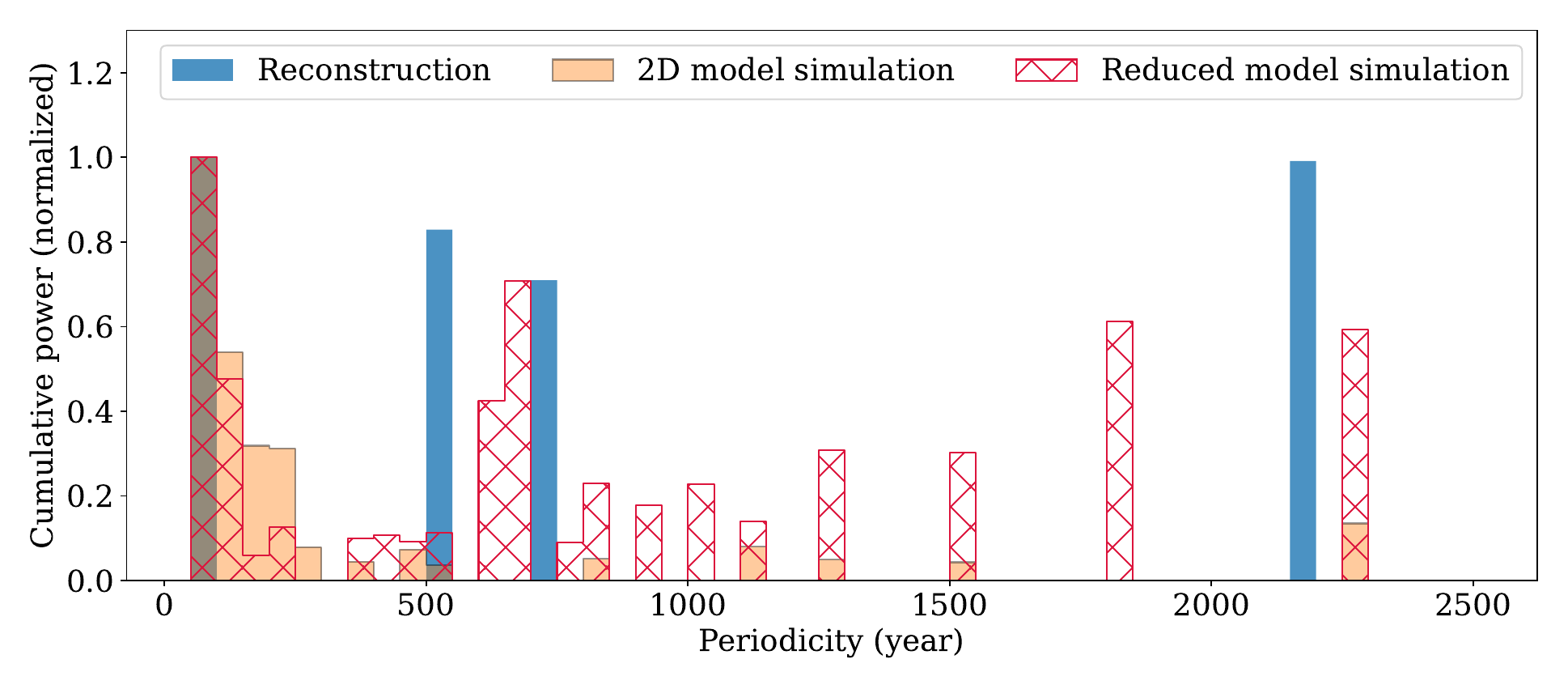}
    \caption{Distributions of cumulative power contained in 50-yr wide bands of periodicities for statistically significant spectral peaks lying above 3-$\sigma$ threshold, in the decadal-averaged time series of (a) reconstructed data (blue), (b) stochastic 2D solar dynamo model simulations (orange), (c) stochastic time delay dynamo model simulations (red cross-hatch). Each of these three distributions is normalized to unity with respect to the bins containing the highest spectral power.}
    \label{fig:sim1}
    \end{center}
\end{figure*}

With the aforementioned setup for the two models, we perform long-term dynamo simulations spanning a hundred millennia producing more than nine thousand solar cycles. All our simulations are far below a chaotic regime and limited to the near critical regime, which is where the current solar dynamo is understood to be operating. In the spatially extended model, we calculate the total unsigned flux of toroidal fields penetrating the meridional plane at 0.677--0.726R$_{\odot}$ in the latitudinal range of 10$^{\circ}$- 45$^{\circ}$ in both hemispheres and consider this as the proxy for the number of sunspots; while in case of the reduced model, the same is represented by the squared toroidal field. For better comparison with the reconstructed solar activity (ref. Fig.\ref{fig:dat1}, top panel), the simulated sunspot number time series is split into an ensemble of 11 consecutive segments, each stretched across a duration of 9,000 years. Fast Fourier Transformation (FFT) of individual epochs in the ensemble reveals the existence of multiple statistically significant periodicities beyond the decadal timescale. It is interesting to note that the longer periodicities denoting supradecadal modulation are not sharply defined, rather they are spread across bands \citep{REIKARD2020JASTP, USOSKIN2023LIVREV}. Histograms in Fig.\ref{fig:sim1} depict the cumulative power of Fourier coefficients for all the epochs, binned into 50-yrs time windows and normalized with respect to the bin having the highest amplitude. The finite spread in the distribution of spectral power can be attributed to the stochastic forcing which also perturbs the cycle durations.

Now, to assess the relative contribution of stochastic perturbation and nonlinearity in driving supersecular solar activity, we completely turn off all stochastic fluctuations in both the dynamo models and simulate 2,000 cycles by keeping every other parameter unchanged as compared to our previous simulations.  Fourier spectra of the resulting time series for both models are shown in  Fig.\ref{fig:sim2}, which clearly indicate the absence of any statistically significant evidence of supradecadal modulation in the simulated solar cycles in the absence of stochastic forcing. Intriguingly, this result remains unchanged for both models even when we increase the order of nonlinear quenching in poloidal sources in both models from quadratic to quartic (see Fig.\ref{fig:sim2}, middle panels). 

The nonlinearities that we incorporate in our models include algebraic $\alpha$-quenching and a buoyancy threshold on the deep seated toroidal field. While other nonlinear mechanisms have been proposed, it is practically impossible to invoke all possible nonlinear mechanisms in any model. We can compensate for this by testing the robustness of our results for super-critical dynamo regime -- by increasing the dynamo number (i.e., the magnitude of the source terms) to sufficiently large values. The method followed to determine the critical $\alpha_{BL}$ for dynamo operation in our models is discussed in Appendix \ref{sec: Appendix}. Results from simulations with varying parameters (effectively with higher dynamo numbers) are shown in the bottom panels of Fig. \ref{fig:sim2}. Notably, in case of the reduced model an increment in $\alpha_{BL}$ makes the fundamental period of the dynamo cycle shift to a larger value \citep{WILMOT2006APJ}. Nevertheless, for both extended and reduced models, there is no excitation of supradecadal modulation beyond the fundamental magnetic cycle in the near-critical dynamo regime. 

\section{Concluding Remarks}

Stellar convection zones host complex processes wherein plasma motions and magnetic fields interact to amplify and sustain large-scale magnetism. On the one hand, nonlinear mechanisms can suppress dynamo action thereby acting as amplitude limiting factors. On the other hand, stochastic forcing acts as an independent source of amplitude fluctuations. As a result, solar-like stars can exhibit a myriad range of activities across different timescales. 

In this work, we address the fundamental question whether non-linearity alone can sustain supradecadal modulation in the near critical dynamo regime -- which independent works suggest to be the current state of operation of the solar dynamo. We emphasize that the presence of stochastic fluctuations extends the range of dynamo operation in our simulations from weakly subcritical to weakly super-critical. Our findings -- based on results obtained from the two complementary dynamo models -- explicitly confirm that nonlinearity is not a sufficient condition and stochastic perturbations are necessary for supradecadal modulation to manifest in the solar cycle. Our findings -- along with independent studies \citep{MININNI2000PRL, MININNI2002PRL, CAMERON2019AANDA} -- reinforce the conclusion that the solar cycle (in its current state) is not chaotically modulated.

An important insight from our simulations is that
supradecadal solar activity modulation are quasi-periodic with no precise periodicities, but rather distributed in narrow bands in the frequency spectrum. Therefore laying emphasis on \textit{exact} periodicities larger than the decadal period (e.g., in Gleissberg cycle, Suess de Vries cycle, etc.) may not be physically meaningful.

We note that our findings are relevant for the interpretation of temporal dynamics associated with a plethora of natural phenomena -- involving a combination of weakly nonlinear, non-deterministic processes -- ranging from solar-stellar interiors to other fluid and MHD systems.

\section{acknowledgements}
     This work is dedicated to the memory of Argentine plasma physicist Daniel Gómez of the Institute of Astronomy and Space Physics and University of Buenos Aires (IAFE, UBA-CONICET). The authors thank Robert Cameron, Ilya Usoskin and Paul Charbonneau for useful discussions, and the anonymous reviewer for their comments. The authors also acknowledge helpful exchanges during the third team meeting of ISSI Team 474 sponsored by the International Space Science Institute, Bern. C.S. acknowledges fellowship from CSIR through grant no. 09/921(0334)/2020-EMR-I. S.M. acknowledges the INSPIRE scholarship (formerly KVPY) from the Department of Science and Technology, Government of India. CESSI is supported by IISER Kolkata, Ministry of Education, Government of India. This research has made use of the VizieR catalogue access tool, CDS, Strasbourg, France.

\section{Data Availability}
We have used reconstructed sunspot number data available at VizieR Online Data Catalog \citep{WU2018VIZIER, 2021yCat..36490141U}. Data from our simulations will be shared upon reasonable request to the corresponding author.

\bibliography{references}{}

\providecommand{\noopsort}[1]{}\providecommand{\singleletter}[1]{#1}%
\begin{thebibliography}{}
\expandafter\ifx\csname natexlab\endcsname\relax\def\natexlab#1{#1}\fi
\providecommand{\url}[1]{\href{#1}{#1}}
\providecommand{\dodoi}[1]{doi:~\href{http://doi.org/#1}{\nolinkurl{#1}}}
\providecommand{\doeprint}[1]{\href{http://ascl.net/#1}{\nolinkurl{http://ascl.net/#1}}}
\providecommand{\doarXiv}[1]{\href{https://arxiv.org/abs/#1}{\nolinkurl{https://arxiv.org/abs/#1}}}

\bibitem[{Allen \& Smith(1996)}]{ALLEN_1996_JSTOR}
Allen, M.~R., \& Smith, L.~A. 1996, Journal of Climate, 9, 3373.
\newblock \url{http://www.jstor.org/stable/26201460}

\bibitem[{Beer {et~al.}(2017)Beer, Tobias, \& Weiss}]{BEER2018MNRAS}
Beer, J., Tobias, S.~M., \& Weiss, N.~O. 2017, Monthly Notices of the Royal Astronomical Society, 473, 1596, \dodoi{10.1093/mnras/stx2337}

\bibitem[{Bhowmik {et~al.}(2023)Bhowmik, Jiang, Upton, Lemerle, \& Nandy}]{Bhowmik2023SSR}
Bhowmik, P., Jiang, J., Upton, L., Lemerle, A., \& Nandy, D. 2023, Space Science Reviews, 219, 40, \dodoi{10.1007/s11214-023-00983-x}

\bibitem[{Biswas {et~al.}(2023)Biswas, Karak, Usoskin, \& Weisshaar}]{BISWAS2023ISSI}
Biswas, A., Karak, B.~B., Usoskin, I., \& Weisshaar, E. 2023, Space Science Reviews, 219, 19, \dodoi{10.1007/s11214-023-00968-w}

\bibitem[{Braun {et~al.}(2005)Braun, Christl, Rahmstorf, Ganopolski, Mangini, Kubatzki, Roth, \& Kromer}]{BRAUN2005NATURE}
Braun, H., Christl, M., Rahmstorf, S., {et~al.} 2005, Nature, 438, 208.
\newblock \url{https://doi.org/10.1038/nature04121}

\bibitem[{Bushby(2006)}]{BUSHBY2006MNRAS}
Bushby, P.~J. 2006, Monthly Notices of the Royal Astronomical Society, 371, 772, \dodoi{10.1111/j.1365-2966.2006.10706.x}

\bibitem[{Cameron \& Schüssler(2015)}]{CAMERON2015SCIENCE}
Cameron, R., \& Schüssler, M. 2015, Science, 347, 1333, \dodoi{10.1126/science.1261470}

\bibitem[{Cameron \& Schüssler(2017)}]{CAMERON2017APJ}
Cameron, R.~H., \& Schüssler, M. 2017, The Astrophysical Journal, 843, 111, \dodoi{10.3847/1538-4357/aa767a}

\bibitem[{Cameron \& Schüssler(2019)}]{CAMERON2019AANDA}
---. 2019, Astronomy and astrophysics, 625, A28, \dodoi{10.1051/0004-6361/201935290}

\bibitem[{Charbonneau(2020)}]{CHARBONNEAU2020LIVREV}
Charbonneau, P. 2020, Living Reviews in Solar Physics, 17, 4, \dodoi{10.1007/s41116-020-00025-6}

\bibitem[{Charbonneau \& Dikpati(2000)}]{Charbonneau2000ApJ}
Charbonneau, P., \& Dikpati, M. 2000, The Astrophysical Journal, 543, 1027, \dodoi{10.1086/317142}

\bibitem[{Chatterjee {et~al.}(2004)Chatterjee, Nandy, \& Choudhuri}]{CHATTERJEE2004AANDA}
Chatterjee, P., Nandy, D., \& Choudhuri, A.~R. 2004, Astronomy \& Astrophysics, 427, 1019.
\newblock \url{https://doi.org/10.1051/0004-6361:20041199}

\bibitem[{{Choudhuri}(1992)}]{Choudhuri1992AandA}
{Choudhuri}, A.~R. 1992, \aap, 253, 277

\bibitem[{Choudhuri \& Karak(2012)}]{CHOUDHURI2012PRL}
Choudhuri, A.~R., \& Karak, B.~B. 2012, Phys. Rev. Lett., 109, 171103, \dodoi{10.1103/PhysRevLett.109.171103}

\bibitem[{{Dasi-Espuig, M.} {et~al.}(2010){Dasi-Espuig, M.}, {Solanki, S. K.}, {Krivova, N. A.}, {Cameron, R.}, \& {Pe\~nuela, T.}}]{DASI2010AANDA}
{Dasi-Espuig, M.}, {Solanki, S. K.}, {Krivova, N. A.}, {Cameron, R.}, \& {Pe\~nuela, T.} 2010, Astronomy and Astrophysics, 518, A7, \dodoi{10.1051/0004-6361/201014301}

\bibitem[{{D'Silva} \& {Choudhuri}(1993)}]{DSILVA1993AANDA}
{D'Silva}, S., \& {Choudhuri}, A.~R. 1993, Astronomy and Astrophysics, 272, 621.
\newblock \url{https://ui.adsabs.harvard.edu/abs/1993AA...272..621D}

\bibitem[{Gupta {et~al.}(2023)Gupta, Basak, \& Nandy}]{Gupta2023ApJ}
Gupta, S., Basak, A., \& Nandy, D. 2023, The Astrophysical Journal, 953, 70, \dodoi{10.3847/1538-4357/acd93b}

\bibitem[{Hathaway(2015)}]{HATHAWAY2015LIVREV}
Hathaway, D.~H. 2015, Living reviews in solar physics, 12, 1.
\newblock \url{https://doi.org/10.1007/lrsp-2015-4}

\bibitem[{Hazra {et~al.}(2023)Hazra, Nandy, Kitchatinov, \& Choudhuri}]{Hazra2023SSR}
Hazra, G., Nandy, D., Kitchatinov, L., \& Choudhuri, A.~R. 2023, Space Science Reviews, 219, 39, \dodoi{10.1007/s11214-023-00982-y}

\bibitem[{Hazra \& Nandy(2019)}]{HAZRA2019MNRAS}
Hazra, S., \& Nandy, D. 2019, Monthly Notices of the Royal Astronomical Society, 489, 4329, \dodoi{10.1093/mnras/stz2476}

\bibitem[{Hazra {et~al.}(2014)Hazra, Passos, \& Nandy}]{HAZRA2014APJ}
Hazra, S., Passos, D., \& Nandy, D. 2014, The Astrophysical Journal, 789, 5.
\newblock \url{https://dx.doi.org/10.1088/0004-637X/789/1/5}

\bibitem[{Jha {et~al.}(2020)Jha, Karak, Mandal, \& Banerjee}]{JHA2020APJL}
Jha, B.~K., Karak, B.~B., Mandal, S., \& Banerjee, D. 2020, The Astrophysical Journal Letters, 889, L19, \dodoi{10.3847/2041-8213/ab665c}

\bibitem[{Jiang(2020)}]{Jiang2020ApJ}
Jiang, J. 2020, The Astrophysical Journal, 900, 19, \dodoi{10.3847/1538-4357/abaa4b}

\bibitem[{{Jouve} {et~al.}(2010){Jouve}, {Proctor, M. R. E.}, \& {Lesur, G.}}]{JOUVE2010AAP}
{Jouve}, {Proctor, M. R. E.}, \& {Lesur, G.} 2010, A\&A, 519, A68, \dodoi{10.1051/0004-6361/201014455}

\bibitem[{Karak(2020)}]{Karak_2020_ApJl_latq}
Karak, B.~B. 2020, The Astrophysical Journal Letters, 901, L35, \dodoi{10.3847/2041-8213/abb93f}

\bibitem[{Karak(2023)}]{KARAK2023LIVREV}
---. 2023, Living Reviews in Solar Physics, 20, 3, \dodoi{10.1007/s41116-023-00037-y}

\bibitem[{Karak \& Miesch(2017)}]{Karak2017APJ}
Karak, B.~B., \& Miesch, M. 2017, The Astrophysical Journal, 847, 69, \dodoi{10.3847/1538-4357/aa8636}

\bibitem[{{Matthaeus} {et~al.}(2007){Matthaeus}, {Breech}, {Dmitruk}, {Bemporad}, {Poletto}, {Velli}, \& {Romoli}}]{MATTHAEUS2007APJ}
{Matthaeus}, W.~H., {Breech}, B., {Dmitruk}, P., {et~al.} 2007, The Astrophysical Journal Letters, 657, L121, \dodoi{10.1086/513075}

\bibitem[{Metcalfe {et~al.}(2016)Metcalfe, Egeland, \& van Saders}]{Metcalfe2016ApJL}
Metcalfe, T.~S., Egeland, R., \& van Saders, J. 2016, The Astrophysical Journal Letters, 826, L2, \dodoi{10.3847/2041-8205/826/1/L2}

\bibitem[{Mininni {et~al.}(2000)Mininni, G\'omez, \& Mindlin}]{MININNI2000PRL}
Mininni, P.~D., G\'omez, D.~O., \& Mindlin, G.~B. 2000, Phys. Rev. Lett., 85, 5476, \dodoi{10.1103/PhysRevLett.85.5476}

\bibitem[{Mininni {et~al.}(2002)Mininni, G\'omez, \& Mindlin}]{MININNI2002PRL}
---. 2002, Phys. Rev. Lett., 89, 061101, \dodoi{10.1103/PhysRevLett.89.061101}

\bibitem[{Miyahara {et~al.}(2004)Miyahara, Masuda, Muraki, Furuzawa, Menjo, \& Nakamura}]{MIYAHARA2004SOLPHYS}
Miyahara, H., Masuda, K., Muraki, Y., {et~al.} 2004, Solar Physics, 224, 317, \dodoi{10.1007/s11207-005-6501-5}

\bibitem[{Nagy {et~al.}(2017)Nagy, Lemerle, Labonville, Petrovay, \& Charbonneau}]{NAGY2017SolPhys}
Nagy, M., Lemerle, A., Labonville, F., Petrovay, K., \& Charbonneau, P. 2017, Solar Physics, 292, 167, \dodoi{10.1007/s11207-017-1194-0}

\bibitem[{Nandy \& Choudhuri(2002)}]{NANDY2002SCIENCE}
Nandy, D., \& Choudhuri, A.~R. 2002, Science, 296, 1671.
\newblock \url{https://doi.org/10.1126/science.1070955}

\bibitem[{Nandy {et~al.}(2023)Nandy, Baruah, Bhowmik, Dash, Gupta, Hazra, Lekshmi, Pal, Pal, Roy, Saha, \& Sinha}]{NANDY2023JASTP}
Nandy, D., Baruah, Y., Bhowmik, P., {et~al.} 2023, Journal of Atmospheric and Solar-Terrestrial Physics, 248, 106081, \dodoi{https://doi.org/10.1016/j.jastp.2023.106081}

\bibitem[{{Nicol} {et~al.}(2009){Nicol}, {Chapman}, \& {Dendy}}]{NICOL2009APJ}
{Nicol}, R.~M., {Chapman}, S.~C., \& {Dendy}, R.~O. 2009, \apj, 703, 2138, \dodoi{10.1088/0004-637X/703/2/2138}

\bibitem[{Olemskoy \& Kitchatinov(2013)}]{Olemskoy2013ApJ}
Olemskoy, S.~V., \& Kitchatinov, L.~L. 2013, The Astrophysical Journal, 777, 71, \dodoi{10.1088/0004-637X/777/1/71}

\bibitem[{Pal {et~al.}(2023)Pal, Bhowmik, Mahajan, \& Nandy}]{PAL2023APJ}
Pal, S., Bhowmik, P., Mahajan, S.~S., \& Nandy, D. 2023, The Astrophysical Journal, 953, 51

\bibitem[{Panchev \& Tsekov(2007)}]{PANCHEV2007JASTP}
Panchev, S., \& Tsekov, M. 2007, Journal of Atmospheric and Solar-Terrestrial Physics, 69, 2391, \dodoi{https://doi.org/10.1016/j.jastp.2007.07.011}

\bibitem[{Passos {et~al.}(2014)Passos, Nandy, Hazra, \& Lopes}]{PASSOS2014AANDA}
Passos, D., Nandy, D., Hazra, S., \& Lopes, I. 2014, Astronomy \& Astrophysics, 563, A18.
\newblock \url{https://doi.org/10.1051/0004-6361/201322635}

\bibitem[{Reikard(2020)}]{REIKARD2020JASTP}
Reikard, G. 2020, Journal of Atmospheric and Solar-Terrestrial Physics, 211, 105465.
\newblock \url{https://www.sciencedirect.com/science/article/pii/S1364682620302686}

\bibitem[{Reinhold {et~al.}(2020)Reinhold, Shapiro, Solanki, Montet, Krivova, Cameron, \& Amazo-Gómez}]{REINHOLD2020SCIENCE}
Reinhold, T., Shapiro, A.~I., Solanki, S.~K., {et~al.} 2020, Science, 368, 518, \dodoi{10.1126/science.aay3821}

\bibitem[{Rempel {et~al.}(2023)Rempel, Bhatia, Bellot Rubio, \& Korpi-Lagg}]{Rempel2023SSR}
Rempel, M., Bhatia, T., Bellot Rubio, L., \& Korpi-Lagg, M.~J. 2023, Space Science Reviews, 219, 36, \dodoi{10.1007/s11214-023-00981-z}

\bibitem[{Saha {et~al.}(2022)Saha, Chandra, \& Nandy}]{SAHA2022MNRASL}
Saha, C., Chandra, S., \& Nandy, D. 2022, Monthly Notices of the Royal Astronomical Society: Letters, 517, L36.
\newblock \url{https://doi.org/10.1093/mnrasl/slac104}

\bibitem[{Scafetta {et~al.}(2016)Scafetta, Milani, Bianchini, \& Ortolani}]{SCAFETTA2016ESR}
Scafetta, N., Milani, F., Bianchini, A., \& Ortolani, S. 2016, Earth-Science Reviews, 162, 24.
\newblock \url{https://www.sciencedirect.com/science/article/pii/S0012825216301453}

\bibitem[{Schrijver {et~al.}(2015)Schrijver, Kauristie, Aylward, Denardini, Gibson, Glover, Gopalswamy, Grande, Hapgood, Heynderickx, Jakowski, Kalegaev, Lapenta, Linker, Liu, Mandrini, Mann, Nagatsuma, Nandy, Obara, {Paul O’Brien}, Onsager, Opgenoorth, Terkildsen, Valladares, \& Vilmer}]{SCHRIJVER2015AdvSR}
Schrijver, C.~J., Kauristie, K., Aylward, A.~D., {et~al.} 2015, Advances in Space Research, 55, 2745, \dodoi{https://doi.org/10.1016/j.asr.2015.03.023}

\bibitem[{Sonett \& Finney(1990)}]{SONETT19990PTRSL}
Sonett, C., \& Finney, S. 1990, Philosophical Transactions of the Royal Society of London. Series A, Mathematical and Physical Sciences, 330, 413.
\newblock \url{https://doi.org/10.1098/rsta.1990.0022}

\bibitem[{Tripathi {et~al.}(2021)Tripathi, Nandy, \& Banerjee}]{TRIPATHI2021MNRASL}
Tripathi, B., Nandy, D., \& Banerjee, S. 2021, Monthly Notices of the Royal Astronomical Society: Letters, 506, L50.
\newblock \url{https://doi.org/10.1093/mnrasl/slab035}

\bibitem[{Usoskin {et~al.}(2016)Usoskin, Gallet, Lopes, Kovaltsov, \& Hulot}]{USOSKIN2016AANDA}
Usoskin, I., Gallet, Y., Lopes, F., Kovaltsov, G., \& Hulot, G. 2016, Astronomy \& Astrophysics, 587, A150.
\newblock \url{https://doi.org/10.1051/0004-6361/201527295}

\bibitem[{Usoskin(2023)}]{USOSKIN2023LIVREV}
Usoskin, I.~G. 2023, Living Reviews in Solar Physics, 20, 2, \dodoi{10.1007/s41116-023-00036-z}

\bibitem[{{Usoskin} {et~al.}(2021){Usoskin}, {Solanki}, {Krivova}, {Hofer}, {Kovaltsov}, {Wacker}, {Brehm}, \& {Kromer}}]{2021yCat..36490141U}
{Usoskin}, I.~G., {Solanki}, S.~K., {Krivova}, N., {et~al.} 2021, VizieR Online Data Catalog, J/A+A/649/A141

\bibitem[{{Usoskin, I. G.} {et~al.}(2021){Usoskin, I. G.}, {Solanki, S. K.}, {Krivova, N. A.}, {Hofer, B.}, {Kovaltsov, G. A.}, {Wacker, L.}, {Brehm, N.}, \& {Kromer, B.}}]{USOSKIN2021AANDA}
{Usoskin, I. G.}, {Solanki, S. K.}, {Krivova, N. A.}, {et~al.} 2021, A\&A, 649, A141, \dodoi{10.1051/0004-6361/202140711}

\bibitem[{van Saders {et~al.}(2016)van Saders, Ceillier, Metcalfe, Aguirre, Pinsonneault, Garc{\'i}a, Mathur, \& Davies}]{vanSaders2016Nature}
van Saders, J.~L., Ceillier, T., Metcalfe, T.~S., {et~al.} 2016, Nature

\bibitem[{Vasiliev \& Dergachev(2002)}]{VASILIEV2002ANNALES}
Vasiliev, S.~S., \& Dergachev, V.~A. 2002, Annales Geophysicae, 20, 115, \dodoi{10.5194/angeo-20-115-2002}

\bibitem[{Weiss \& Tobias(2015)}]{WEISS2016MNRAS}
Weiss, N.~O., \& Tobias, S.~M. 2015, Monthly Notices of the Royal Astronomical Society, 456, 2654, \dodoi{10.1093/mnras/stv2769}

\bibitem[{Wilmot-Smith {et~al.}(2006)Wilmot-Smith, Nandy, Hornig, \& Martens}]{WILMOT2006APJ}
Wilmot-Smith, A., Nandy, D., Hornig, G., \& Martens, P. 2006, The Astrophysical Journal, 652, 696.
\newblock \url{https://dx.doi.org/10.1086/508013}

\bibitem[{Wilmot-Smith {et~al.}(2005)Wilmot-Smith, Martens, Nandy, Priest, \& Tobias}]{WILMOT2005MNRAS}
Wilmot-Smith, A.~L., Martens, P. C.~H., Nandy, D., Priest, E.~R., \& Tobias, S.~M. 2005, Monthly Notices of the Royal Astronomical Society, 363, 1167, \dodoi{10.1111/j.1365-2966.2005.09514.x}

\bibitem[{{Wu} {et~al.}(2018){Wu}, {Usoskin}, {Krivova}, {Kovaltsov}, {Baroni}, {Bard}, \& {Solanki}}]{WU2018VIZIER}
{Wu}, C.~J., {Usoskin}, I.~G., {Krivova}, N., {et~al.} 2018, VizieR Online Data Catalog, J/A+A/615/A93.
\newblock \url{https://ui.adsabs.harvard.edu/abs/2018yCat..36150093W}

\bibitem[{Yeates {et~al.}(2025)Yeates, Bertello, Pevtsov, \& Pevtsov}]{Yeates2025ApJ}
Yeates, A.~R., Bertello, L., Pevtsov, A.~A., \& Pevtsov, A.~A. 2025, The Astrophysical Journal, 978, 147, \dodoi{10.3847/1538-4357/ad99d0}

\bibitem[{Yeates {et~al.}(2008)Yeates, Nandy, \& Mackay}]{YEATES2008APJ}
Yeates, A.~R., Nandy, D., \& Mackay, D.~H. 2008, The Astrophysical Journal, 673, 544, \dodoi{10.1086/524352}

\end{thebibliography}
\bibliographystyle{aasjournal}

\appendix

\section{Determination of critical alpha parameters}
\label{sec: Appendix}

In order to determine the critical regime of dynamo operation, we calculate the mean toroidal flux in both spatially extended and reduced models and plot its variation with the non-stochastic forcing of $\alpha_{BL}$. In 2D model, the toroidal flux is calculated near the base of the convective zone in a latitudinal extent corresponding to the active latitude belt. Whereas, in the reduced model, squared magnitude of the toroidal magnetic field component is considered as a proxy for the toroidal flux. 

The parameter regime where the dynamo response just exceeds the threshold for buoyant eruption in the 2D BL dynamo model is considered the critical regime; for this model, the magnitude of $\alpha_{BL}$ for which this is achieved in our model is 21 ms$^{-1}$ (see the left panel of Fig.\ref{fig:app1}). Equivalently, 
for the reduced time-delay dynamo model the critical value of $\alpha_{BL}$ is 0.21 yr$^{-1}$  (see the right panel of Fig.\ref{fig:app1}).

\begin{figure*}
    \begin{center}    
    \includegraphics[width=0.9\textwidth]{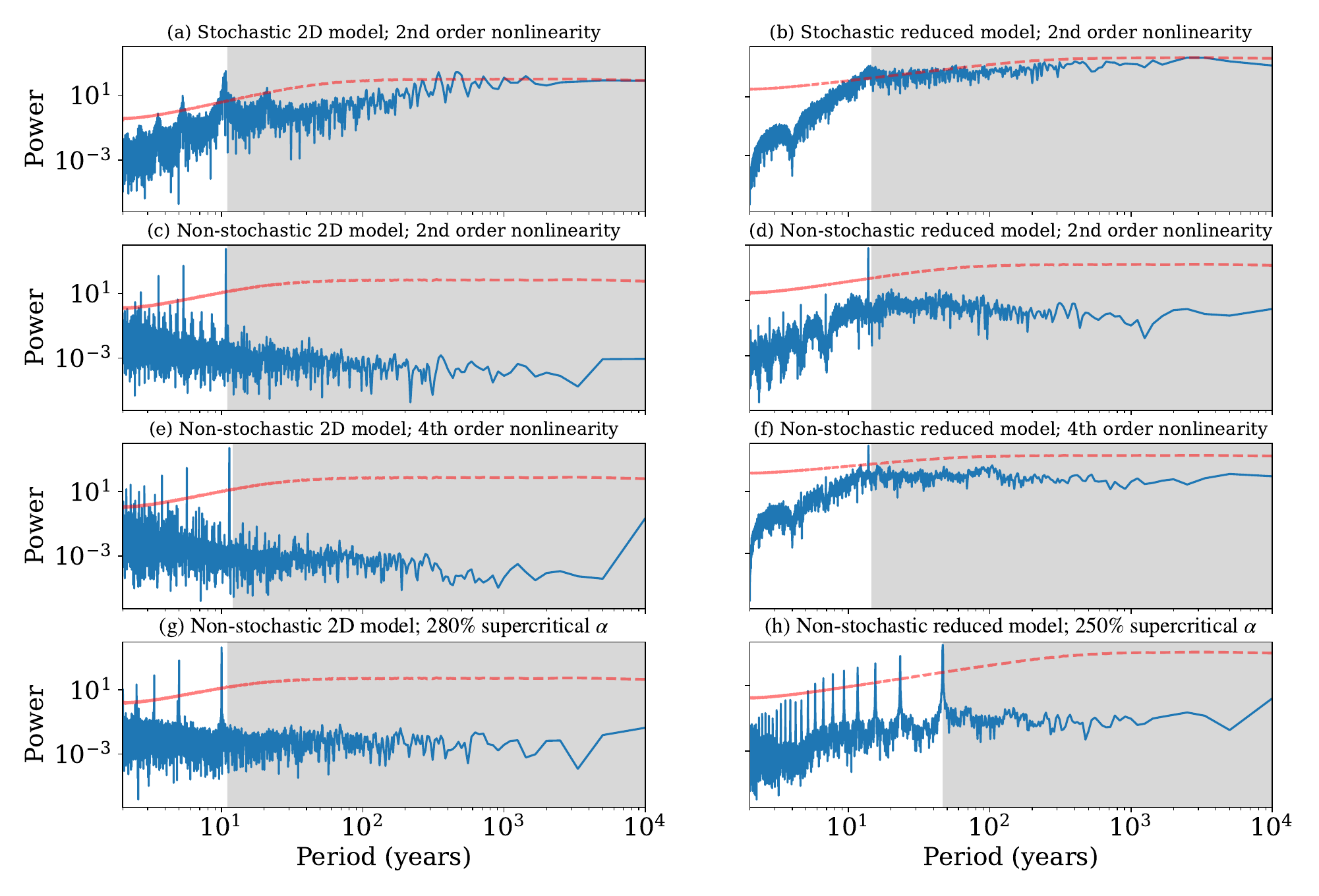}
    \caption{Power spectrum of simulated sunspot time series without any stochastic fluctuation in the poloidal sources. Results are from 2D solar dynamo model simulations (panel a, c \& e) and time delay dynamo model simulations (panel b, d \& f).  Top panels (a \& b) show results for  2$^{nd}$ order nonlinearity incorporated in the poloidal sources, whereas, middle panels are similar to top panels, but with the order of nonlinearity, portrayed in Eqs.\eqref{eq:q1} and \eqref{eq:q2}, increased to four. Bottom panels demonstrate the results for highly supercritical regime of dynamo operation. The red-dashed curve denotes the statistical 3-$\sigma$ threshold. The gray-shaded region in all the panels denote the periodicity regime that correspond to supradecadal modulation. In all of these non-stochastic simulations, there is no trace of statistically significant spectral peak in the supradecadal regime. Refer to the appendix for the exact amplitudes of poloidal sources used in these simulations.}
    \label{fig:sim2}
    \end{center}
\end{figure*}

\begin{figure}
    \begin{center}    
    \includegraphics[width=\textwidth]{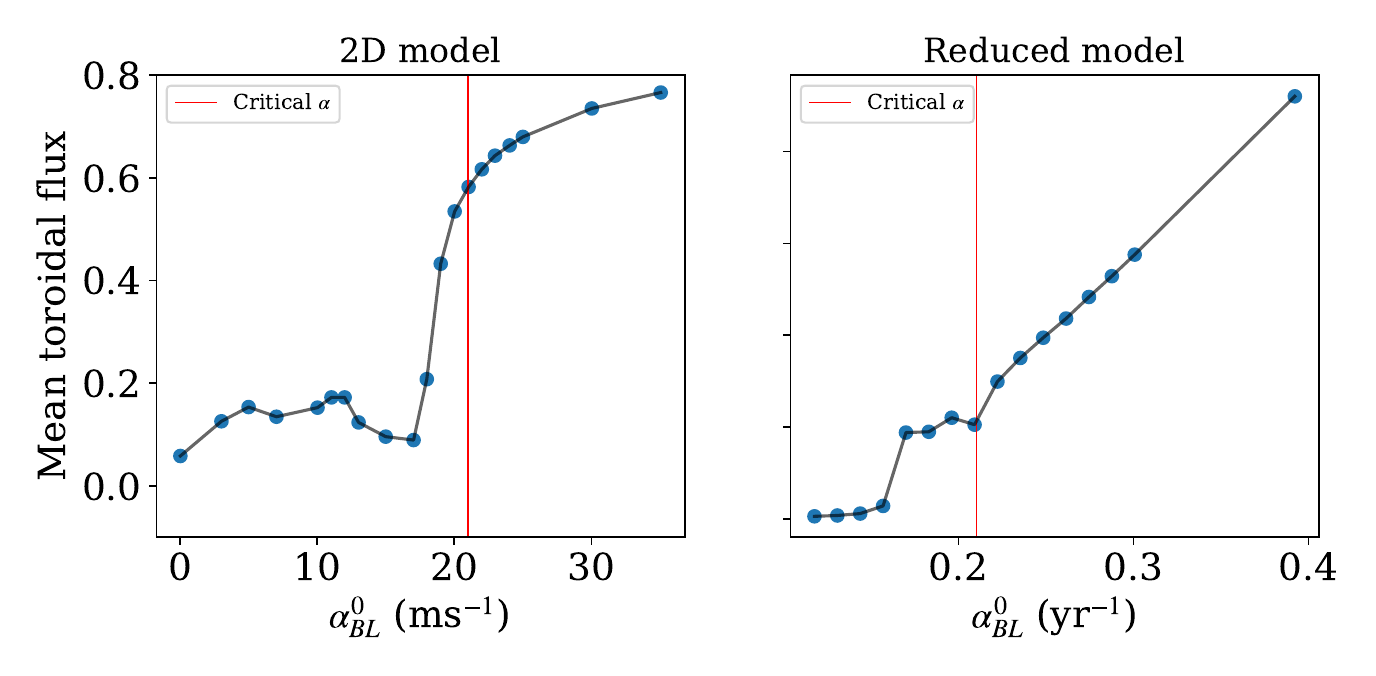}
    \caption{Variation of the simulated mean toroidal flux with $\alpha_{BL}$ in the absence of stochastic forcing, for the 2D model (left panel) and the reduced model (right panel).}
    \label{fig:app1}
    \end{center}
\end{figure}

Our simulations (except the last row in Fig.\ref{fig:sim2}) are carried out in a near-critical regime, slightly above the critical limit. The time delay dynamo operates at 1.4 times the critical alpha, whereas, in our 2D model, $\alpha_{BL}$ is set at 1.2 times the critical limit.

\end{document}